\documentclass[aps,showpacs,twocolumn,prl]{revtex4-1}
\usepackage{graphicx,color,epsfig,rotate}
\usepackage{times,xspace}
\usepackage{color}
\usepackage{graphicx}
\usepackage{dcolumn}
\usepackage{amssymb}
\usepackage{amsmath}
\usepackage{amsfonts}
\usepackage{docs}%
\usepackage{bm}%
\usepackage{wrapfig}
\usepackage[colorlinks=true,linkcolor=blue,pagecolor=blue,filecolor=blue,menucolor=blue,urlcolor=blue,citecolor=blue,anchorcolor=blue]{hyperref}%
\usepackage{bm}
\usepackage{graphics}
\usepackage{graphicx}
\usepackage{color}
\usepackage{amssymb}
\usepackage{standalone}
\usepackage[toc,page]{appendix}
\usepackage{import}
\usepackage{enumerate}
\usepackage{sidecap}

\newcommand{\bsigma}{{\boldsymbol \sigma}}
\newcommand{\bnabla}{{\boldsymbol \nabla}}

\newcommand{\bbeta}{{\boldsymbol \eta}}
\usepackage{amssymb,amsmath,bm}

\begin{document}


\title{Josephson Junction through a Disordered Topological Insulator with Helical Magnetization}


\author{Alexander Zyuzin}
\author{Mohammad Alidoust }
\author{Daniel Loss}
\affiliation{Department of Physics,
University of Basel, Klingelbergstrasse 82, CH-4056 Basel,
Switzerland}
\date{\today}

\begin{abstract}

We study supercurrent and proximity vortices in a Josephson junction made of disordered surface states of a three-dimensional topological insulator with a proximity induced in-plane helical magnetization. In a regime where the rotation period of helical magnetization is larger than the junction width, we find supercurrent $0$-$\pi$ crossovers as a function of junction thickness, magnetization strength, and parameters inherent to the helical modulation and surface states. The supercurrent reversals are associated with proximity induced vortices, nucleated along the junction width, where the number of vortices and their locations can be manipulated by means of the superconducting phase difference and the parameters mentioned above. 
\end{abstract}


\maketitle

\section{Introduction}%
Topological insulators (TIs) form a new state of matter which has offered novel prospects towards topological superconducting spintronics and topological quantum computation \cite{pankratov, kane_rmp,zhang_rmp}. The TIs provide a platform to observe quantum relativistic phenomena, stemming from strong spin orbit interaction, such as the quantum spin Hall effect, spin-momentum locking, and manipulation of Dirac fermions (see reviews \cite{kane_rmp,zhang_rmp}).
The surface of three-dimensional TI in the presence of time reversal symmetry hosts metallic helical states, {\it i.e.}, for each momentum on
the Fermi surface of the surface states, the spin has a rigidly defined direction, transverse to the momentum. 
Interestingly, the interplay of superconductivity and magnetism at the surface of a TI
may give rise to topological superconductivity and Majorana fermions which has been receiving strong interest both experimentally
and theoretically \cite{Volovik,exp_ti2a, exp_ti2b, exp_ti2c,exp_ti2c2,exp_ti4a,exp_ti4b,exp_ti4c}. 

To the best of our knowledge, the theoretical studies on superconducting TI structures have considered fully ballistic surface states. 
However, experimentally realistic systems inevitably contain impurity scattering processes 
that can play a key role in the actual quantum transport through the surface states \cite{exp_ti4a,burkov1,Schwab,exp_ti2b,Kurter_TI_exp}. A theory for the disordered limit of `{\it superconducting}' TI with the possibility of inclusion of `{\it magnetism}' is still lacking. Therefore, with the rapid experimental progress in TI heterostructures such a theoretical framework is becoming more essential for describing and predicting important physical phenomena in this field.  
The quasiclassical formalism is capable of providing such a theoretical technique \cite{brgt_rmp1,brgt_rmp2,brgt_rmp3}.

\begin{figure}[t!]
\includegraphics[width=8.0cm,height=4.50cm]{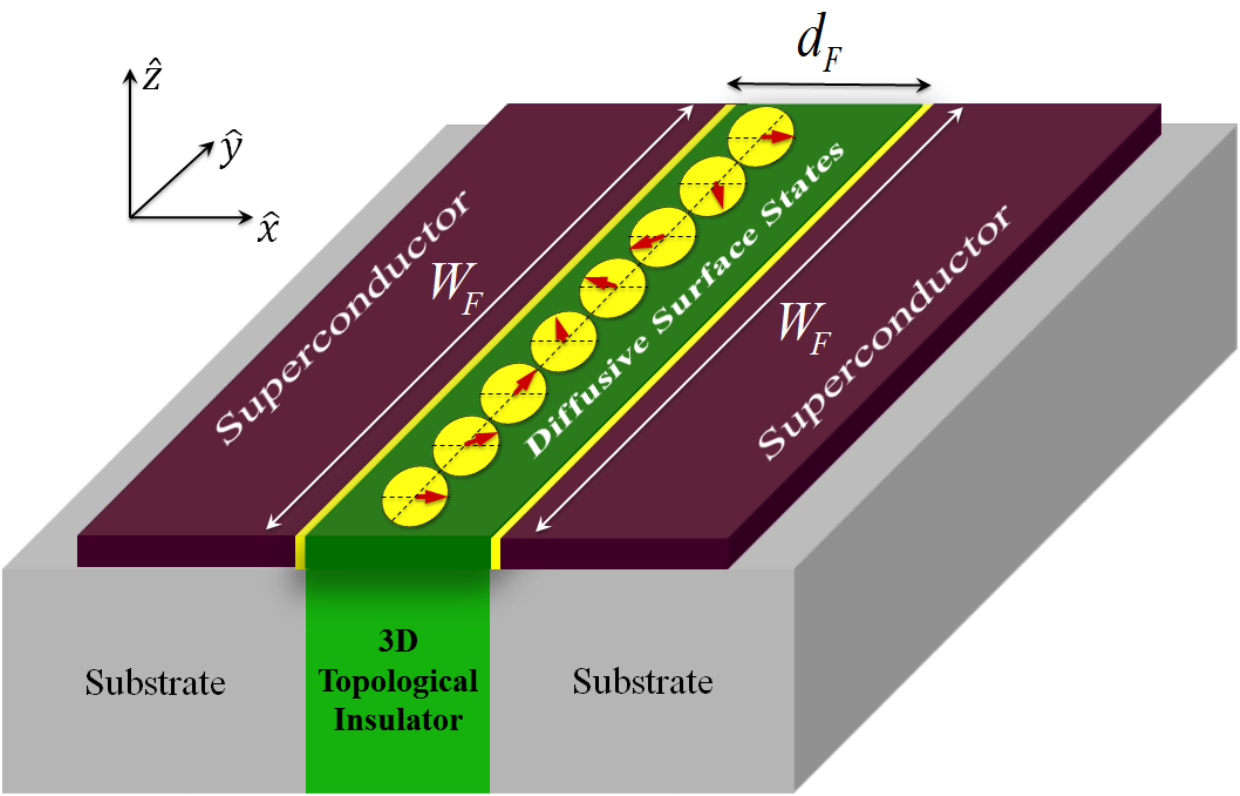}
\caption{\label{fig:model} (Color online). 
Schematic of a superconductor (S) - topological insulator (TI) - superconductor junction with a helical magnetization pattern. The magnetization vector follows a helical pattern given by $\mathbf{h}(\mathbf{ r})=h_0(\cos Qy,\sin Qy,0)$. The junction plane resides in the $xy$ plane so that the S-TI interfaces lie in the $y$ direction at $x=0, d_F$. The junction has a length and width of $d_F$ and $W_F$, respectively. The superconducting electrodes are
connected to the  diffusive surface states of the TI through tunneling barriers.
 }
\end{figure}

It is well understood that the ground state of a uniformly magnetized 3D metallic superconductor - ferromagnet - superconductor (SFS) junction can be reverted from $0$ to $\pi$ by varying the magnetization strength, F layer thickness, and the system temperature, for a review see Refs.~\cite{brgt_rmp1,brgt_rmp2,brgt_rmp3}. Such a $0$-$\pi$ transition, first predicted theoretically \cite{buz_0_pi_0a,buz_0_pi_0b,buz_0_pi_0c}, is now well established experimentally \cite{Ryazanov_0a,Ryazanov_0b,Ryazanov_0c}. The $0$-$\pi$ crossover is important for applications in quantum computations and for the development of ultrafast switches in superconducting spintronics and functional nanodevices \cite{brgt_rmp1,brgt_rmp2,brgt_rmp3}. On the other hand, it is known that the free energy of the SFS Josephson junction in the presence of the spin-orbit interaction in the F region has a minimum at phase difference $\phi = \phi_0,~(\neq 0, \pi)$, \cite{phi_jncta,phi_jnctb,phi_jnctc,phi_jnctd,phi_jncte} (the so-called $\phi_0$-junctions). Signatures of the Josephson $\phi_0$-junction based on a nanowire quantum dot were recently observed experimentally in Ref. \cite{phi0exp}. We also note that a $\phi_0$ junction may be also achieved in a different situation where the superconducting phase is random along the junction width \cite{zyuzin}.

It was shown that the critical supercurrent shows no reversal in Josephson junctions made of ballistic surface states of a TI (S-TI-S) with proximity induced `uniform' magnetization \cite{tanaka_sfs}. 
This follows from the fact that the TI surface states do not respond to a static and uniform in-plane magnetic (or exchange) field due to the momentum-spin locking of Dirac fermions. In-plane magnetic fields via Zeeman effect simply shift the surface Dirac cones in momentum space. Thus, the current-phase relation has a phase shift of $\phi_0$, proportional to the junction thickness and the Zeeman energy.

Here we generalize a quasiclassical approach for disordered 
surface states of three-dimensional TIs in the presence of $s$-wave superconductivity and in-plane magnetization with an arbitrary pattern.
We derive the Eilenberger equation \cite{eiln} which describes the system from a fully ballistic to a weakly disordered regime. We also derive the Usadel equation \cite{usadel}, governing the transport of quasiparticles in fully disordered TI surface states where diffusive motion dominates the ballistic one. 

We first employ the Usadel approach for a S-TI-S junction with uniform magnetization and show that similarly to the ballistic limit the critical supercurrent is featureless and an exchange field parallel to the S-TI interfaces transfers into the superconducting phase difference across the junction.
We then study the supercurrent in a diffusive 2D S-TI-S with an in-plane helical magnetization, illustrated in Fig. \ref{fig:model}. 
Interestingly, we find that multiple supercurrent reversals and proximity vortices can appear by simply manipulating the helical magnetization parameters and varying the junction thickness.

The rest of the article is organized as follows. In the next section (Sec. \ref{R_and_D}) we describe our model and discuss main results. In Sec. \ref{sec:derivation} we present the detailed analytical derivations of the Eilenberger and Usadel equations as well as the supercurrent through TI, and finally in Sec. \ref{Conclusion} we present concluding remarks.

\section{ Main Results }\label{R_and_D}%
The equation for the Green function of quasiparticle at the surface of 3D TI reads:
\begin{eqnarray}\nonumber \label{Green01}
\left( \begin{array}{cc}
-i\omega_n+\hat{H}(\mathbf{ r})& 0 \\
0 & i\omega_n +\hat{\sigma}^y\hat{H}^*(\mathbf{ r})\hat{\sigma}^y
\end{array} \right)\check{G}(\omega_n;\mathbf{ r},\mathbf{ r}')
\\
=\delta(\mathbf{ r}-\mathbf{ r}')+\frac{1}{\pi \nu \tau} \check{G}(\omega_n;\mathbf{ r},\mathbf{ r}) \check{G}(\omega_n;\mathbf{ r},\mathbf{ r}'),
\end{eqnarray}
where the Hamiltonian $\hat{H}(\mathbf{r}) =-i\alpha (\bnabla \times \mathbf{ e}_z)\cdot\hat{\bm \sigma} +\mathbf{ h}(\mathbf{r}) \cdot\hat{\bsigma} -\mu$ describes the
surface states of the TI with a proximity induced in-plane exchange field $\mathbf{ h}(\mathbf{r})=(h_x(\mathbf{r}), h_y(\mathbf{r}), 0)$. Here, $\alpha$ is the Fermi velocity, $\mu$ is the chemical potential,
$\mathbf{e}_z$ is an unit vector normal to the surface of TI, and $\hat{\bm \sigma}$ is a vector comprised of the Pauli matrices acting on the spin degree of freedom. We assume $\hbar=k_B=1$ throughout the paper.  
The Green function is averaged over a nonmagnetic scattering potential $V(\mathbf{ r})$, which is assumed Gaussian $\langle V(\mathbf{ r})V(\mathbf{ r}')\rangle = \delta(\mathbf{ r}-\mathbf{ r}')/\pi \nu \tau$, where $\tau$ is the mean free time of particles in the disordered system and $\nu = \mu/2\pi \alpha^2$ is the density of states per spin at the Fermi level of the normal state of the TI.
The matrix structure of the Green function in the rotated particle-hole and spin basis has the following form:
\begin{eqnarray}\label{Green00}
\check{G}(\omega_n;\mathbf{ r},\mathbf{ r}')=\left( \begin{array}{cc}
-\hat{G}(\omega_n;\mathbf{ r},\mathbf{ r}')& -\hat{F}(\omega_n;\mathbf{ r},\mathbf{ r}')i\hat{\sigma}^y \\\nonumber
-i\hat{\sigma}^y\hat{F}^{\dag}(\omega_n;\mathbf{ r},\mathbf{ r}')& \hat{\sigma}^y\hat{\bar{G}}(\omega_n;\mathbf{ r},\mathbf{ r}')\hat{\sigma}^y
\end{array} \right),
\end{eqnarray}
where  $\omega_n =\pi T(2n+1)$ is the Matsubara frequency, $T$ is the temperature, and $n \in \mathbb{Z}$. The check symbol $\check{1}$ represents $4\times 4$ matrices in the particle-hole and spin spaces, while the hat symbol $\hat{1}$ defines $2\times2$ matrices. Notice that the off-diagonal components of the Green function describe the penetration of Cooper pairs into the surface states.

In order to solve Eq. (\ref{Green01}) we employ the quasiclassical approximation which results in the Eilenberger equation  \cite{eiln}
for the surface channels:
\begin{eqnarray}\label{Eilenberger0}
\frac{\alpha}{2}\Big\{\hat{\bbeta}, \bnabla \check{g}\Big\}=\bigg[\check{g} , \omega_n \hat{\tau}^z+i\mathbf{ h}\cdot\hat{\bsigma}\hat{\tau}^z+i \mu\hat{\bbeta} \cdot\mathbf{ n}_F+\frac{\langle \check{g} \rangle}{ \tau}\bigg],~~~~
\end{eqnarray}
where we have performed a Fourier transformation of the Green function Eq. (\ref{Green00}) with respect to the relative space arguments and then taken the integral over $\xi_p = \alpha p -\mu$ which results in $\check{g}(\omega_n; \mathbf{ R}, \mathbf{ n}_F) = \int \frac{d\xi_p}{\pi i} \check{G}(\omega_n; \mathbf{ R}, \mathbf{ p})$. Here, $\mathbf{n}_F = \mathbf{p}_F/|\mathbf{p}_F|$ is an unit vector in the direction of momentum $\mathbf{p}_F$ at Fermi level $\mu = \alpha p_F$, $\hat{\bbeta} \equiv (-\sigma^y,\sigma^x)$ is the vector of Pauli matrices and $\hat{\tau}^z$ is a Pauli matrix acting in the particle-hole space.
The disorder potential scatter quasiparticles in random directions in the momentum space. Therefore, one can integrate the quasiclassical Green function over all possible directions of quasiparticle momentum $\langle \check{g} \rangle$.
To find a solution to Eq. (\ref{Eilenberger0}), we expand the Green function through Pauli matrices $\check{g}= (\hat{g}'\hat{1} + \hat{\mathbf{ g}}'\cdot\hat{\bbeta}+\hat{ g}'_z\hat{\sigma}^z)/2$, where $\hat{\mathbf{ g}}' = (\hat{g}'_x,\hat{g}'_y, 0)$.
The spin structure of $\check{g}$ in the limit of $|\mathbf{h}| \ll \mu$ is defined by the conduction band projector $(1+\hat{\bbeta} \cdot\mathbf{ n}_F)/2$. We find that the main contributing components of $\check{g}$ commute with $\hat{\eta} \mathbf{n}_F$. Hence, the 
$\hat{g}'_{z}$ component is smaller than 
$\hat{g}'$ and $\hat{g}'_{x,y}$ by a factor of $\sim\mathrm{max}(\frac{1}{\tau}, |\mathbf{ h}|)/\mu \ll 1$ and can be neglected.   

\begin{SCfigure*}
  \centering
\includegraphics[width=12.80cm,height=5.40cm]{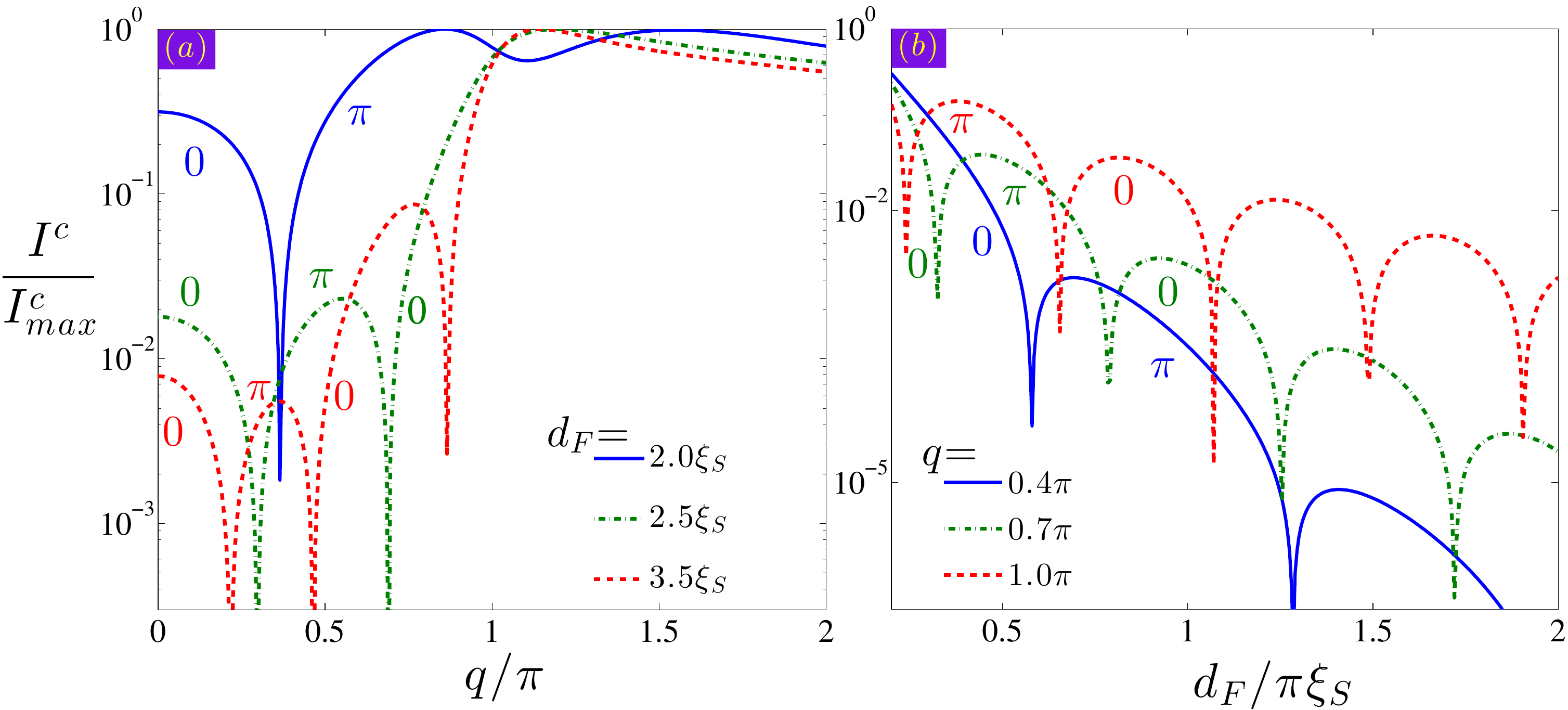}
\caption{\label{fig:Ic_vs_q} (Color online). Normalized critical supercurrent through diffusive TI surface states
with an induced helical magnetization. Panel ($a$) shows the critical current as a function of rotation parameter $q$ at three different values of junction length: $d_F=2.0\xi_S, 2.5\xi_S, 3.5\xi_S$. Panel ($b$) shows the critical current as a function of junction length $d_F$ for $q=0.1\pi, 0.4\pi$, and $0.7\pi$. The junction ground state oscillates between $0$ and $\pi$ superconducting phase difference by varying $q, d_F, h_0$. 
 }
\end{SCfigure*}

In the diffusive limit the Green function can be expanded up to the first two terms of 2D harmonics, namely $\hat{g}' (\omega_n; \mathbf{ R},\mathbf{ n}_F) = \hat{g}_s(\omega_n; \mathbf{ R}) + \mathbf{ n}_F\cdot \hat{\mathbf{ g}}_a (\omega_n; \mathbf{ R})$, 
where the zero harmonic is isotropic and its amplitude is larger than the first harmonic. 
We substitute this expansion into Eq. (\ref{Eilenberger0}) and perform an integration over momentum directions. By taking a spin trace we arrive at $\hat{\mathbf{g}}_a = -2\alpha \tau \hat{g}_s \hat{\bnabla} \hat{g}_s$, which results in the Usadel equation \cite{usadel}:
\begin{eqnarray}\label{Usadel-Maintext}
D \hat{\bnabla}(\hat{g}_s \hat{\bnabla} \hat{g}_s) =  [\omega_n\hat{\tau}^z, \hat{g}_s].
\end{eqnarray}
Here, $D=\alpha^2 \tau$ denotes the diffusion coefficient in the TI. We have defined a covariant derivative $\hat{\bnabla}$ so that $\hat{\bnabla}X = \bnabla X + \frac{i}{\alpha}(h_x \mathbf{ e}_y -h_y \mathbf{ e}_x)[\hat{\tau}^z, X]$,
in which $\mathbf{e}_{x,y}$ are unit vectors in the $x,y$ directions. Contrary to the zero harmonics of the Green function expansion, the first harmonic satisfies a non-diffusive equation due to the fast spin relaxation time proportional to the mean free time at the surface. This finding is consistent with that of the electron spin and charge densities at the TI surface  in the absence of superconductivity \cite{burkov1, Schwab}. We note that the Usadel equation for an electron gas with spin orbit coupling in the presence of a Zeeman field was derived in Refs. \onlinecite{huzet_bergeret}. Here we however focus on the disordered surface states of a topological insulator which are described by the Dirac Hamiltonian. We utilize the derived Usadel equation to study the supercurrent and proximity induced vortices' profile in a 2D Josephson junction with helical magnetization.

Using the definition of quasiclassical Green functions, we obtain an expression for the charge current density in the diffusive limit: 
\begin{equation}\label{currentD0}
\mathbf{ J} = -\frac{\pi i \sigma_N}{4 e} T\sum_n \mathrm{Tr} [\hat{\tau}^z \hat{g}_s \hat{\bnabla} \hat{g}_s],
\end{equation}
where $\sigma_N =2 e^2 \nu D$ is the normal-state conductivity. To calculate the charge supercurrent flow across the junction shown in Fig. \ref{fig:model}, one needs to find a solution to Eq. (\ref{Usadel-Maintext}) in the TI region and match the Green function at S-TI boundaries. Here, we assume that the TI surface is sandwiched between two spin-orbit free $s$-wave superconducting electrodes. We will consider the limit of low transparency of the S-TI interfaces so that the spin-singlet Cooper pairs tunnel from the superconducting leads into the surface states and thus employ the Kupriyanov-Lukichev boundary conditions for the Green function at the contacts \cite{boundary_c1,boundary_c2}.  The leakage intensity of the superconducting correlations is controlled by the parameter $\gamma \gg 1$, which is the ratio of resistance per unit area of the surface of the tunneling barrier to the resistivity of the diffusive TI  surface states. To derive the boundary conditions for the S-TI-S junction considered in this paper, one should follow Refs. \cite{boundary_c1,boundary_c2} and consider the effect of momentum-spin locking. 
One therefore straightforwardly arrives at the following expression for the boundary conditions: $2\gamma \hat{g}_s \mathbf{ n}\cdot\hat{\bnabla} \hat{g}_s = [\hat{g}_s, \hat{g}_{\text{SC}}]$,
where the unit vector $\mathbf{ n}$ points normal to the boundary and the Green function in the superconducting lead $\hat{g}_{\text{SC}}(\omega_n)$ is given by its bulk solution.

To begin, we first consider a wide S-TI-S junction with a uniform in-plane exchange field: $\mathbf{h}(\mathbf{ R})=(h_x, h_y, 0)$, where the junction width is larger than its length $W_F\gg d_F$, as shown in Fig. \ref{fig:model}. 
Deriving the Green function and plugging into the expression for the charge current density Eq. (\ref{currentD0}), using $I=\int_{-W_F/2}^{+W_F/2} dy J_x(y)$, we obtain:
\begin{eqnarray}\label{Current}\nonumber
&&I = \frac{\pi}{2e}\frac{d_F^2}{\gamma^2 R_N} {\cal N} \sin\left(\phi +\frac{2h_yd_F}{\alpha}\right),\\
&&{\cal N} = T\sum_n \frac{|\Delta|^2}{\omega_n^2+|\Delta|^2}\frac{\mathrm{csch}(k_{n}d_F)}{k_{n}d_F},
\end{eqnarray}
where $R_N ={d_F}/{\sigma_N W_F}$ 
, $|\Delta|$ and $\phi$ denote the superconducting gap and phase difference across the contact, and $k_{n}=\sqrt{2|\omega_n|/D + (2h_x/\alpha)^2}$. 
As seen, the $h_x$ component of the magnetization, along the current flow direction, plays a depairing role and supresses the supercurrent monotonically. The transverse component 
$h_y$, however, causes a shift in the superconducting phase difference. Similar effects were also predicted in the fully ballistic S-TI-S counterparts via the Bogoliubov de Gennes approach \cite{tanaka_sfs}. Therefore, one concludes that the nonmagnetic disorders in a uniformly magnetized S-TI-S heterostructure are unable to alter the phase shift in the current phase relation and only modify
the amplitude of the critical current through the surface states of TI, the same as temperature.

\begin{figure}[t!]
\includegraphics[width=7.10cm,height=4.0cm]{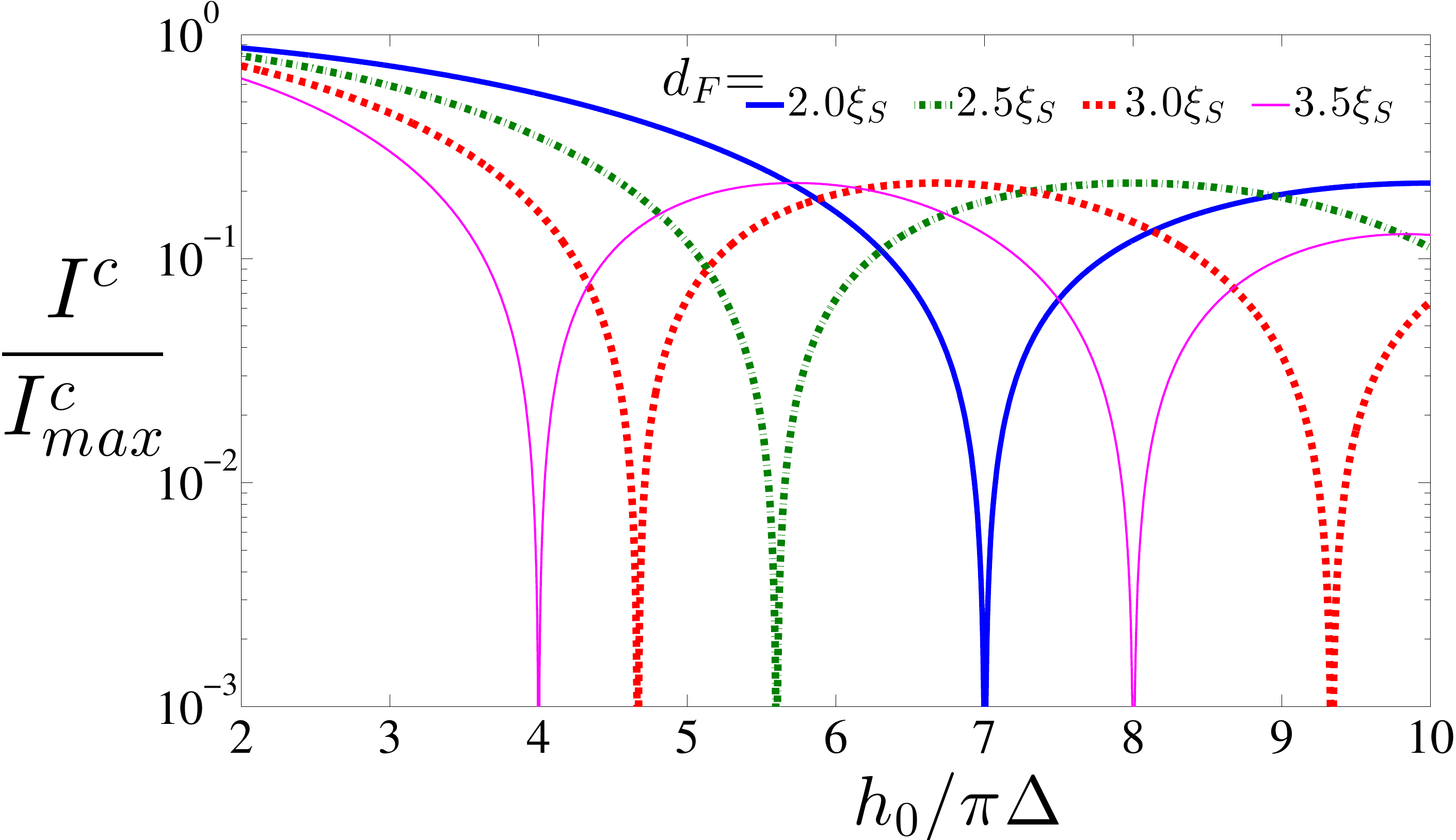}
\caption{\label{fig:fnrhfr} (Color online). Normalized critical supercurrent  through the TI surface channel with a slow rotating magnetization, $q=0.3\pi$, as a function of magnetization intensity $h_0$ for different values for the junction thickness $d_F= 2.0\xi_S, 2.5\xi_S, 3.0\xi_S, 3.5\xi_S$.  
}
\end{figure}

We now turn to a 2D S-TI-S junction with helical magnetization depicted in Fig. \ref{fig:model}. The helical pattern is given by $\mathbf{h}(\mathbf{ R})=h_0(\cos Qy,\sin Qy,0)$, where $Q=q/W_F$ and $q$ determines the actual pattern of the magnetization. The helical magnetization with a period of $\approx 10$nm was already observed experimentally in manganese on a tungsten substrate through spin-polarized tunneling experiments \cite{chrl_mg_exp}.  
In order to solve Eq. (\ref{Usadel-Maintext}) we consider limits $W_F\gg d_F$ for $q\ll 1$ and $W_F\gg q d_F$ for $q\gg 1$.
We substitute the Green function, obtained for the configuration shown in Fig. \ref{fig:model}, into Eq. (\ref{currentD0}) and find the supercurrent: 
\begin{eqnarray}\label{crnt_cycl}\nonumber
&&I = \frac{\pi}{2e}\frac{d_F^2}{\gamma^2 R_N} \int_{-W_F/2}^{+W_F/2} \frac{dy}{W_F}  {\cal N}_{y} \sin\bigg(\phi+ \frac{2 h_0d_F}{\alpha} \sin Qy\bigg),\\
&&{\cal N}_{y} = T\sum_n \frac{|\Delta|^2}{\omega_n^2+|\Delta|^2}\frac{\mathrm{csch}(k_{n,y}d_F)}{k_{n,y}d_F},
\end{eqnarray}
where $k_{n,y} =\sqrt{2|\omega_n|/D + (2h_0/\alpha)^2 \cos^2 Qy}$.
As seen, the integrand in Eq. (\ref{crnt_cycl}) is a highly nonlinear function of $y$.
We have numerically integrated Eq. (\ref{crnt_cycl}) and plotted the critical supercurrent as a function of $q$ and $d_F$ in Fig. \ref{fig:Ic_vs_q}. To plot the currents, we have normalized lengths by a length scale $\xi_S=\sqrt{D/|\Delta|}$, energies by the superconducting gap $|\Delta|$, defined the Thouless energy $\varepsilon_T=D/d_F^2$, and normalized the critical current $I^c$ by its maximum value $I^c_{max}$ in the interval calculated. 
We consider a low temperature regime and set a fixed normalized temperature at $T/T_c=0.01$, where $T_c$ is the superconducting critical temperature. 
Fig. \ref{fig:Ic_vs_q}a illustrates the critical supercurrent profile as a function of $q$ at $h_0=5.0 \pi|\Delta|$ 
for different values of $d_F$. At small values of junction length, $d_F\approx 2.0 \xi_S$, the supercurrent changes sign at a single value of $0.4\pi<q<0.5\pi$. At larger values of junction length, however, we see that the supercurrent undergoes multiple reversals and the Josephson ground state oscillates between $0$ and $\pi$ phase differences. Fig. \ref{fig:Ic_vs_q}b shows the critical current as a function the junction length for different $q$. These patterns of critical current are reminiscent of those found in Ref. \onlinecite{eschrig} for a 3D metallic superconductor-ferromagnet-superconductor junction. One should note, however, that the nature of the $0$-$\pi$ transitions explored in this paper is essentially different from those of Ref. \onlinecite{eschrig}. Here, the supercurrent reversals appear solely due to the helical magnetization which is locked to the quasiparticle momentum. Whereas in a metallic system the exchange field plays a dephasing role on the Cooper pairs and causes oscillations in the Cooper pair amplitude which yields supercurrent reversal. To gain more insights, we may simplify Eq. (\ref{crnt_cycl}) by considering a slow rotating magnetization so that $q< 1$ and neglect the $x$ component of magnetization. In this limit, we find a simple expression for the supercurrent which clearly illustrates the damped oscillatory behaviour of critical current as a function of $qh_0d_F/\alpha$:
\begin{eqnarray}\label{cc_frnhfr}
I = \frac{\pi}{2e}\frac{d_F^2}{\gamma^2 R_N} {\cal N}_0\frac{\sin(qh_0d_F/\alpha)}{qh_0d_F/\alpha} \sin\phi.
\end{eqnarray}
The normalized critical supercurrent $I^c$ as a function of $h_0$ for various values of $d_F$ is shown in Fig. \ref{fig:fnrhfr}. It is clear that the critical supercurrent changes sign upon varying the magnetization intensity $h_0$. 
Considering Eq. (\ref{cc_frnhfr}), the supercurrent $0$-$\pi$ crossovers appear at $qh_0d_F/\alpha=n\pi$ ($n=\pm 1, \pm 2, ...$). 
Taking $\alpha\approx 10^7$ cm/s and $h_0 \approx 5$ meV, the first $0$-$\pi$ transition happens in a junction of length $d_F=200$ nm for $q\approx 0.2\pi$.

In addition to the current density 
the absolute value of the Cooper pair wave function \cite{brgt_rmp3} can be expressed in the slow rotation mode by:
\begin{eqnarray}\label{cur_dnsy}
&&U(y) = \frac{T}{4\gamma}\sum_n\frac{|\Delta|\text{csch}(k_nd_F/2)}{\sqrt{\omega_n^2+|\Delta|^2}}\cos\bigg(\frac{\phi}{2}+\frac{h_0d_F}{\alpha}Qy\bigg),~~~~
\end{eqnarray}
which can provide information about the proximity vortices. It is evident that both the current density and $U(y)$ vanish at locations $y/W_F=\frac{\phi\pm n\pi}{2qh_0d_F}\alpha$, $n=\pm 1, \pm 3, ...$ provided that $-1/2<y/W_F<+1/2$. These points correspond to the normal cores of proximity vortices in TI \cite{nphys_2015_m}. It is straightforward to show that the proximity vortices are present in the case of helical magnetization and $q$ can control the number of vortices. 

\section{Derivations of main results}\label{sec:derivation}

\subsection{Eilenberger Equation}
The Hamiltonian describing
the Rashba type surface states of a three-dimensional topological insulator (TI) with the proximity induced ferromagnetism with an in-plane exchange field, $\mathbf{h}(\mathbf{ r})=(h_x(\mathbf{ r}), h_y(\mathbf{ r}), 0)$, reads: 
\begin{equation}\label{eq:haml_1}
\hat{H}(\mathbf{r}) = 
-i\alpha (\bnabla \times \mathbf{ e}_z)\cdot\hat{\bm \sigma} +\mathbf{h}(\mathbf{ r}) \cdot\hat{\bsigma} -\mu
\end{equation}
where $\alpha$ is the Fermi velocity, characterizing the surface states,
$\mathbf{e}_z$ is a unit vector normal to the surface of TI with chemical potential $\mu$, $\hat{\bm \sigma}$ is a vector comprised of the Pauli matrices, and the hat symbol ($\hat{1}$) denotes $2\times 2$ matrices. We here consider the Rashba type throughout our calculations. However, through the available symmetries in the formulations, our results can be easily extended to the Dresselhaus type, $-i\beta_D\bnabla\cdot\hat{\bsigma}$, by simple prescriptions given at the end of our Sec. \ref{sec:usa}. Therefore, without lossing generality, we derive the Eilenberger and Usadel equations for the Rashba type surface states.

The electron Green function is defined by:
\begin{subequations}\label{GF_comps}
\begin{eqnarray}
G_{ab}(\tau_1,\tau_2; \mathbf{ r}_1, \mathbf{ r}_2) &=& - \langle T_{\tau} \Psi_{a} (\tau_1,\mathbf{ r}_2) \Psi_{b}^{\dag}(\tau_2,\mathbf{ r}_2)  \rangle,~~~~~~~\\
\bar{G}_{ab}(\tau_1,\tau_2; \mathbf{ r}_1, \mathbf{ r}_2) &=& - \langle T_{\tau} \Psi^{\dag}_{a} (\tau_1,\mathbf{ r}_1) \Psi_{b}(\tau_2,\mathbf{ r}_2)  \rangle,~~~~~~~\\
F_{ab}(\tau_1,\tau_2; \mathbf{ r}_1, \mathbf{ r}_2) &=& + \langle T_{\tau} \Psi_{a} (\tau_1,\mathbf{ r}_1) \Psi_{b}(\tau_2,\mathbf{ r}_2)  \rangle,~~~~~~~\\
F^{\dag}_{ab}(\tau_1,\tau_2; \mathbf{ r}_1, \mathbf{ r}_2) &=& + \langle T_{\tau} \Psi^{\dag}_{a} (\tau_1,\mathbf{ r}_1) \Psi^{\dag}_{b}(\tau_2,\mathbf{ r}_2)  \rangle,~~~~~~~
\end{eqnarray}
\end{subequations}
where $a\equiv\uparrow,\downarrow$ and $b\equiv\uparrow,\downarrow$ define the electron spin projections, 
and $\tau_1, \tau_2$ are the imaginary times at $\mathbf{ r}_1, \mathbf{ r}_2$, respectively.
To simplify our derivations, we have introduced a unitary rotation in the particle-hole and spin spaces: 
\begin{eqnarray}\label{Green0}
\check{G}(\omega_n;\mathbf{ r},\mathbf{ r}')=\left( \begin{array}{cc}
-\hat{G}(\omega_n;\mathbf{ r},\mathbf{ r}')& -\hat{F}(\omega_n;\mathbf{ r},\mathbf{ r}')i\hat{\sigma}^y \\\nonumber
-i\hat{\sigma}^y\hat{F}^{\dag}(\omega_n;\mathbf{ r},\mathbf{ r}')& \hat{\sigma}^y\hat{\bar{G}}(\omega_n;\mathbf{ r},\mathbf{ r}')\hat{\sigma}^y
\end{array} \right),\\
\end{eqnarray}
here we performed a Fourier transformation $\check{G}(\omega_n;\mathbf{ r},\mathbf{ r}') = T\sum_n e^{-i\omega_n\tau} \check{G}(\tau;\mathbf{ r},\mathbf{ r}')$ to Matsubara frequency $\omega_n =\pi T(2n+1)$ where $T$ is the temperature and $n \in \mathbb{Z}$.
Check symbol $\check{1}$ defines $4\times 4$ matrices in the particle-hole and spin spaces. The off diagonal components of the Green function (\ref{Green0}) matches to those of superconducting bulk solutions at the boundaries and describe the penetration of the Cooper pairs into the TI surface states.

Considering a Gaussian distribution for the nonmagnetic impurity scattering potential $V(\mathbf{ r})$, one finds 
\begin{equation}
\langle V(\mathbf{ r})V(\mathbf{ r}')\rangle = \frac{1}{\pi \nu \tau}\delta(\mathbf{ r}-\mathbf{ r}'),\;\;\;\nu = \frac{\mu}{2\pi \alpha^2}
\end{equation}
where $\tau$ is the mean free time of particles in the disordered system and $\nu$ is the density of states per spin at the Fermi level of the normal state of TI. In this particle-hole and spin rotated system, we arrive at the following equation for the particle Green function averaged over the scattering potential:
\begin{eqnarray}\nonumber \label{Green1}
\left( \begin{array}{cc}
-i\omega_n+\hat{H}(\mathbf{ r})& -\epsilon_g \\
\epsilon_g^* & i\omega_n +\hat{\sigma}^y\hat{H}^*(\mathbf{ r})\hat{\sigma}^y
\end{array} \right)\check{G}(\omega_n;\mathbf{ r},\mathbf{ r}')
\\
=\delta(\mathbf{ r}-\mathbf{ r}')+\frac{1}{\pi \nu \tau} \check{G}(\omega_n;\mathbf{ r},\mathbf{ r}) \check{G}(\omega_n;\mathbf{ r},\mathbf{ r}'),
\end{eqnarray}
the second term on the right hand side of Eq. (\ref{Green1}) is defined by the self-energy part of the averaged Green function. Here we allow for the proximity induced minigap $\epsilon_g$
in the surface states of the TI.

We first subtract from Eq. (\ref{Green1}) its conjugated equation and perform a Fourier transformation with respect to the relative space arguments:
\begin{equation}\label{Green2}
\check{G}(\omega_n; \mathbf{ R} + \frac{\delta\mathbf{ r}}{2}, \mathbf{ R} - \frac{\delta\mathbf{ r}}{2})= \int\frac{d\mathbf{ p}}{(2\pi)^2} \check{G}(\omega_n; \mathbf{ R}, \mathbf{ p}) e^{i \mathbf{ p}\cdot \delta \mathbf{ r}},
\end{equation}
in which we have defined $\mathbf{ R}= (\mathbf{ r}+\mathbf{ r}')/2$, $\delta \mathbf{ r} = \mathbf{ r}-\mathbf{ r}'$, and denoted the momentum vector of the quasiparticles by $\mathbf{ p}$. We also define a new parameter $\xi_p= \alpha p - \mu$ and by making use of the fact that the Green function peaks at the Fermi surface, we obtain the quasiclassical Green function 
\begin{eqnarray}\label{G2}
\check{g}(\omega_n; \mathbf{ R}, \mathbf{ n}_F) = \int \frac{d\xi_p}{\pi i} \check{G}(\omega_n; \mathbf{ R}, \mathbf{ p}).
\end{eqnarray}
Using this assumption, we obtain the Eilenberger equation, \cite{eiln}:
\begin{eqnarray}\label{Interm}\nonumber
\frac{\alpha}{2}\{\hat{\bbeta}, \bnabla \check{g}\}=\bigg[\check{g} , \omega_n \hat{\tau}^z+i\mathbf{ h}\cdot\hat{\bsigma}\hat{\tau}^z+i \mu\hat{\bbeta}\cdot \mathbf{ n}_F+i\check{\epsilon}_g+\frac{\langle \check{g} \rangle}{ \tau}\bigg],\\
\end{eqnarray}
where $\hat{\bm \bbeta} = (-\hat{\sigma}^y,\hat{\sigma}^x)$, $\check{\epsilon}_g = \hat{\sigma}^0(-\epsilon_g\hat{\tau}^{+} +\epsilon_g^*\hat{\tau}^-)/2$, and $\hat{\tau}^{\pm} =\hat{\tau}^x\pm i \hat{\tau}^y$. This equation governs moving quasiparticles in ballistic (where $1/\tau\rightarrow 0$) systems and those with weak nonmagnetic impurities with a finite $\tau$.

\subsection{Usadel Equation}\label{sec:usa}
The Eilenberger equation can be simplified in systems with strong disorders such that $\mu> 1/\tau > |\omega_n|, |h| ,|\epsilon_g|$. In this case, the quasiparticles follow diffusive trajectories which is the so-called diffusive regime \cite{usadel}. 
In the diffusive regime of the surface states, one can integrate the quasiclassical Green function, Eq. (\ref{G2}), over all possible directions of quasiparticles' momentum:
\begin{eqnarray}
\langle \check{g}(\omega_n; \mathbf{ R}, \mathbf{ n}_F) \rangle \equiv \int_0^{2\pi}\frac{d\Omega_{\mathbf{ n}_F}}{2\pi}\check{g}(\omega_n; \mathbf{ R}, \mathbf{ n}_F), \;\mathbf{ n}_F =\frac{\mathbf{ p}_F}{|\mathbf{ p}_F|}, ~~~~~\;\;
\end{eqnarray}
where $\Omega_{\mathbf{ n}_F}$ is the polar angle of the vector $\mathbf{n}_F$ in the plane of TI surface.
We note that Eq. (\ref{Interm}) is similar to the kinetic equation for the Keldysh component of the Green function of TI in the normal state \cite{Schwab}.

To find a solution to Eq. (\ref{Interm}), we expand the Green function through the Pauli matrices:
\begin{equation}
\check{g}= \frac{\hat{g}'\hat{1} + \hat{\mathbf{ g}}' \cdot\hat{\bbeta}+\hat{ g}'_z\hat{\sigma}^z}{2},
\end{equation}
where $\hat{\mathbf{ g}}' = (\hat{g}'_x,\hat{g}'_y, 0)$.
We assume that the Fermi energy in the conduction band of the TI is much larger than all other energy scales available in the system. We immediately find that leading contributions to $\check{g}$ commute with $\hat{\eta} \mathbf{n}_F$. Thus, $\hat{g}'_z$ component is smaller than 
$\hat{g}'$ and $\hat{g}'_{x,y}$ by a factor of $\sim\mathrm{max}(\frac{1}{\tau}, |\mathbf{ h}|)/\mu \ll 1$ and can be neglected. Also, the spin structure of $\check{g}$ is proportional to the conduction band projector $(1+\hat{\bbeta} \mathbf{ n}_F)/2$ in the limit of $|\mathbf{h}| \ll \mu$. We then set $\hat{\mathbf{g}}'=\mathbf{n}_F \hat{g}'$ and propose a solution to Eq. (\ref{Interm}) in the form of a direct product of two $2 \times 2$ matrices:
\begin{eqnarray}\label{anzats}
\check{g}(\omega_n; \mathbf{ R},\mathbf{ n}_F) = \hat{g}'(\omega_n; \mathbf{ R},  \mathbf{ n}_F) 
\frac{1+ \hat{\bbeta} \cdot\mathbf{ n}_F}{2},
\end{eqnarray}
where we define
\begin{eqnarray}\hat{g}' =\left( \begin{array}{cc}
-g& -if\\
if^{*}&\bar{g}
\end{array} \right).
\end{eqnarray}
Notice that the conduction band projector satisfies: $(1+\hat{\bbeta}\mathbf{ n}_F)^2/4 = (1+\hat{\bbeta} \mathbf{ n}_F)/2$ and  
serves as the source of superconducting triplet correlations in the system. Substituting 
expression (\ref{anzats}) into Eq. (\ref{Green1}), we arrive at the following equation:  
\begin{eqnarray}\nonumber\label{Eilenberger}
\alpha (\hat{\bbeta}+ \mathbf{ n}_F)\cdot
\bnabla\hat{g}' &=&\bigg[\hat{g}'
(1+\hat{\bbeta} \cdot\mathbf{ n}_F), \omega_n\hat{\tau}^z + i\mathbf{ h}\cdot\hat{\bsigma}
\hat{\tau}^z\\
&+&i\check{\Delta}
+\frac{\langle  \hat{g}'
(1+\hat{\bbeta} \cdot\mathbf{ n}_F) \rangle}{2 \tau} \bigg].~~~~
\end{eqnarray}
In the limit of diffusive motion of quasiparticles, the Green function can be expanded through the first two terms of 2D harmonics, namely:
\begin{equation}\label{expansion}
\hat{g}' (\omega_n; \mathbf{ R},\mathbf{ n}_F) = \hat{g}_s(\omega_n; \mathbf{ R}) + \mathbf{ n}_F \cdot \hat{\mathbf{ g}}_a (\omega_n; \mathbf{ R}),
\end{equation}
where the zero harmonic in the expansion is isotropic and its amplitude is much larger than the first harmonic: $\hat{g}_s \gg \mathbf{ n}_F \cdot \hat{\mathbf{ g}}_a$. 

We now substitute the expanded Green function, Eq. (\ref{expansion}), into Eq. (\ref{Eilenberger}) and perform an integration over momentum directions. By taking a spin trace we finally arrive at:
\begin{eqnarray}\label{Int1}
\alpha \hat{\bnabla}\cdot \hat{\mathbf{g}}_a = 2 [\hat{g}_s,\omega_n\hat{\tau}^z+i\check{\epsilon}_g].
\end{eqnarray}
Here, we have defined a new operator $\hat{\bnabla}$ so that  
\begin{equation}
\hat{\bnabla}X = \bnabla X+ \frac{i}{\alpha}(h_x \mathbf{ e}_y -h_y \mathbf{ e}_x)[\hat{\tau}^z, X],
\end{equation}
where $\mathbf{e}_{x,y}$ are unit vectors in the $x,y$ directions.
Multiplying Eq. (\ref{Eilenberger}) by $\mathbf{ n}_F$ and integrating it over momentum directions, 
we find the following expression for the first harmonic term of the Green function expansion $\hat{\mathbf{g}}_a$:
\begin{eqnarray}\label{Int2}
 \hat{\mathbf{g}}_a = -2\alpha \tau \hat{g}_s \hat{\bnabla} \hat{g}_s.
\end{eqnarray}
To obtain the Usadel equation, it suffices we substitute Eq. (\ref{Int2}) into Eq. (\ref{Int1}) which yields:
\begin{eqnarray}\label{Usadel}
D \hat{\bnabla}\cdot(\hat{g}_s \hat{\bnabla} \hat{g}_s) =  [\omega_n\hat{\tau}^z+i\check{\epsilon}_g, \hat{g}_s],
\end{eqnarray}
where the diffusion coefficient is denoted by $D=\alpha^2 \tau$. In the case $\check{\epsilon}_g=0$, Eq. \ref{Usadel} leads to Eq. \ref{Usadel-Maintext}.
The singlet part of the Green function, $\hat{g}_s$, satisfies the Usadel equation Eq. (\ref{Usadel}). In contrast, due to fast spin relaxation time which is given by the mean free time in the TI, the motion of the spin part, $\hat{\mathbf{g}}_a$, is not diffusive, and satisfies Eq. (\ref{Int2}). 
To obtain the Eilenberger and Usadel equations for surface states with the Dresselhaus type of spin-orbit coupling, one simply should use ${\bm \eta}=(\sigma_x,\sigma_y)$ in the Eilenberger equation, Eq. (\ref{Interm}), and change $\alpha$ to $\beta_D$. In the Usadel equation, Eq. (\ref{Usadel}), however one can perform the replacement below:
\begin{equation}
\hat{\bnabla} X= \bnabla X+ \frac{i}{\beta_D}(h_x \mathbf{ e}_x + h_y \mathbf{ e}_y)[\hat{\tau}^z, X].
\end{equation}

\subsection{Boundary Conditions and the Current Density}

We consider low transparency limit of the interface (tunneling barrier) between the TI and superconductor \cite{boundary_c1,boundary_c2}.
We neglect the inverse proximity effect so that the Green function in the superconductor at the interface is given by its bulk solution. Assuming that the electron tunneling across the TI-S interface is spin-conserving, while it is not momentum conserving, we eventually arrive at the following expression
\begin{equation}\label{BC_sup}
2\gamma \hat{g}_s \mathbf{ n}\cdot\hat{\bnabla} \hat{g}_s = [\hat{g}_s, \hat{g}_{\text{SC}}],
\end{equation}
where $\mathbf{ n} $ is the unit vector normal to the boundary and
\begin{equation}
\hat{g}_{\text{SC}}(\omega_n) = \frac{1}{\sqrt{\omega_n^2 + |\Delta|^2}}[ \omega_n \hat{\tau}^z + \frac{i}{2} (-\Delta \hat{\tau}^+ + \Delta^* \hat{\tau}^-)],
\end{equation}
is the Green function in the superconductor, in which $|\Delta|$ is the superconducting gap. The low transparency of the interface leads to large parameter $\gamma \gg 1$, which is the ratio of resistance per unit area of the surface of the tunneling barrier to the resistivity of the TI.

Let us now present the current density in the TI.
The current density flowing across the surface of TI is given by: 
\begin{equation}\label{crnt3}
\mathbf{ J}= \frac{e\alpha}{2} T \sum_n \lim_{\mathbf{ r}'\rightarrow \mathbf{ r}} \mathrm{Tr}[\hat{\tau}^z \hat{\bbeta} \check{G}(\omega_n, \mathbf{ r},\mathbf{ r}')],
\end{equation} 
where $e>0$ is the absolute value of electron charge. Rewriting the current density through the quasiclassical Green function, we find: 
\begin{equation}\label{crnt_2}
\mathbf{ J}  = i\pi \frac{e\alpha}{4}T \sum_n \mathrm{Tr} [\hat{\tau}^z \hat{\mathbf{ g}}_a].
\end{equation}
Substituting expression (\ref{Int2}) into (\ref{crnt3}), we obtain the current density in the diffusive limit, Eq.~(\ref{currentD0}).\\

\subsection{Derivation of Supercurrent through Topological Insulator with Uniform and Helical Magnetizations}

In the low proximity limit, we can expand the Green function around the bulk solution which yields:
\begin{eqnarray}\hat{g}_s(\omega_n;\mathbf{ R})=\left( \begin{array}{cc}
\mathrm{sign}(\omega_n)& -i{f_+}(\omega_n;\mathbf{ R}) \\
 i{f_-}(\omega_n;\mathbf{ R})&-\mathrm{sign}(\omega_n)
\end{array} \right).
\end{eqnarray}
The low proximity limit is experimentally relevant and can be easily achieved in temperatures near the superconducting critical temperature or low transparent SC-TI contacts, for instance.
To derive the current density, and consequently the total supercurrent
through a helical magnetization, we use the Usadel equation,
(\ref{Usadel}). We will consider homogeneous and helical magnetization
as shown in Fig. 1 of main text, which can be described by:
\begin{subequations}\label{eq:mg}
\begin{eqnarray}
&&\mathbf{h}=(h_x,h_y,0),\\
&&\mathbf{h}( y)=h_0(\cos Qy,\sin Qy,0),
\end{eqnarray} 
\end{subequations}
where $Q=q/W_F$. 
To avoid complication, we assume that the inverse proximity effect is neglegible and the magnetization is restricted within $0<x<d_F$. Moreover, we consider a situation where the Josephson penetration length $\lambda_J$ is larger than the junction width and ignore the effect of magnetic field induced by the supercurrent itself \cite{zagoskin,birge}. Otherwise, one should solve the Usadel equation together with the Maxwell equations self-consistently.
We assume that the junction width ($W_F$) is larger than the length ($d_F$), $W_F\gg d_F$ and 
arrive at the following differential equations for the anomalous components of the Green function: 
\begin{eqnarray}\label{eq:chir_Us}
\left(\partial_x \mp \frac{2ih_y(y)}{\alpha}\right)^2 f_{\pm} - \frac{4h^2_x(y)}{\alpha^2}f_{\pm}=\frac{2|\omega_n|}{D} f_{\pm}.
\end{eqnarray}
The corresponding boundary conditions in this case result in:
\begin{subequations}\label{gf_chiral}
\begin{eqnarray}
&&\gamma[\partial_x \mp 2ih_y(y)/\alpha]f_{\mp}|_{x=-d_F/2}=\frac{|\Delta|e^{\mp i\phi/2}}{\sqrt{\omega_n^2+|\Delta|^2}},~~~~~~\\
&&\gamma[\partial_x \mp 2ih_y(y)/\alpha]f_{\mp}|_{x=+d_F/2}=-\frac{|\Delta|e^{\pm i\phi/2}}{\sqrt{\omega_n^2+|\Delta|^2}},~~~~~~
\end{eqnarray}
\end{subequations}
where $\phi$ is the superconducting phase difference across the junction and the transparency of SC-TI contacts can be controlled through parameter $\gamma$. To derive boundary conditions (\ref{gf_chiral}), we assume that $\Delta \gg |\omega_n| f_{+}$, which is justified by the SC-TI interface with low transparency.
The Usadel equation and associated boundary conditions yeild the following solutions:
\begin{widetext}
\begin{subequations}\label{sol_chir}
\begin{eqnarray}
f_{+} &=&\frac{-|\Delta|}{\gamma k_{n,y} \sqrt{\omega_n^2 +|\Delta|^2}}\bigg[ \frac{\mathrm{ch}(k_{n,y}(x-d_F/2))}{\mathrm{sh}(k_{n,y}d_F)}e^{i\phi/2+i\frac{2h_y(y)}{\alpha}(x+d_F/2)}
+ \frac{\mathrm{ch}(k_{n,y}(x+d_F/2))}{\mathrm{sh}(k_{n,y}d_F)} e^{-i\phi/2+i\frac{2h_y(y)}{\alpha}(x-d_F/2)} \bigg],~~~~~~\\
f_{-} &=&\frac{-|\Delta|}{\gamma k_{n,y} \sqrt{\omega_n^2 +|\Delta|^2}}\bigg[ \frac{\mathrm{ch}(k_{n,y}(x-d_F/2))}{\mathrm{sh}(k_{n,y}d_F)}e^{-i\phi/2-i\frac{2h_y(y)}{\alpha}(x+d_F/2)}
+ \frac{\mathrm{ch}(k_{n,y}(x+d_F/2))}{\mathrm{sh}(k_{n,y}d_F)} e^{i\phi/2-i\frac{2h_y(y)}{\alpha}(x-d_F/2)} \bigg],~~~~~~
\end{eqnarray}
\end{subequations}
\end{widetext}
where we have introduced a wave-vector 
\begin{equation}
k_{n,y}=\sqrt{2|\omega_n|/D + (2h_x(y)/\alpha)^2}.
\end{equation}
The supercurrent density in the low proximity limit we consider here involves the anomalous components of Green function:
\begin{eqnarray}\label{crnt2}\nonumber
J_x(y) &=& -\frac{\pi i \sigma_N}{4e} T\sum_n \Big\{ f_+(\partial_x +2ih_y(y)/\alpha)f_{-}\\
&-& f_-(\partial_x-2ih_y(y)/\alpha)f_{+}\Big\}.
\end{eqnarray}

To find the total current flow across the junction in the $x$ direction, one needs to integrate the current density, given by expression (\ref{crnt2}), over the junction width, $W_F$, along the $y$ axis: $I=\int_{-W_F/2}^{+W_F/2} dy J_x(y)$. By plugging the solutions found, (\ref{sol_chir}), into (\ref{crnt2}), we obtain the total charge supercurrent, flowing through the junction;
\begin{eqnarray}\nonumber\label{cur_sup}
&&I = \frac{\pi}{2e}\frac{d_F^2}{\gamma^2 R_N} \int_{-W_F/2}^{+W_F/2} \frac{dy}{W_F}  {\cal N}_y \sin\bigg(\phi+ \frac{2 h_y(y)d_F}{\alpha}\bigg),\\
&&{\cal N}_y = T\sum_n \frac{|\Delta|^2}{\omega_n^2+|\Delta|^2}\frac{\mathrm{csch}(k_{n,y}d_F)}{k_{n,y}d_F}.
\end{eqnarray}
It is instructive to consider several limiting cases of expression (\ref{cur_sup}). In the limit of homogeneous exchange field we immediately obtain Eq. (\ref{Current}).
Also, by considering the slow rotation limit, where $q< 1$, of the helical magnetization given by Eq. (\ref{eq:mg}) we find  Eq. (\ref{cc_frnhfr}).
As seen, the critical Josephson current in this slow rotating regime, $q<1$, shows similar features, as a function of $qh_0d_F/\alpha$, to those of conventional metallic wide junctions subject to an external magnetic field. 
  
\begin{figure}[b!]
\includegraphics[width=8.50cm,height=7.50cm]{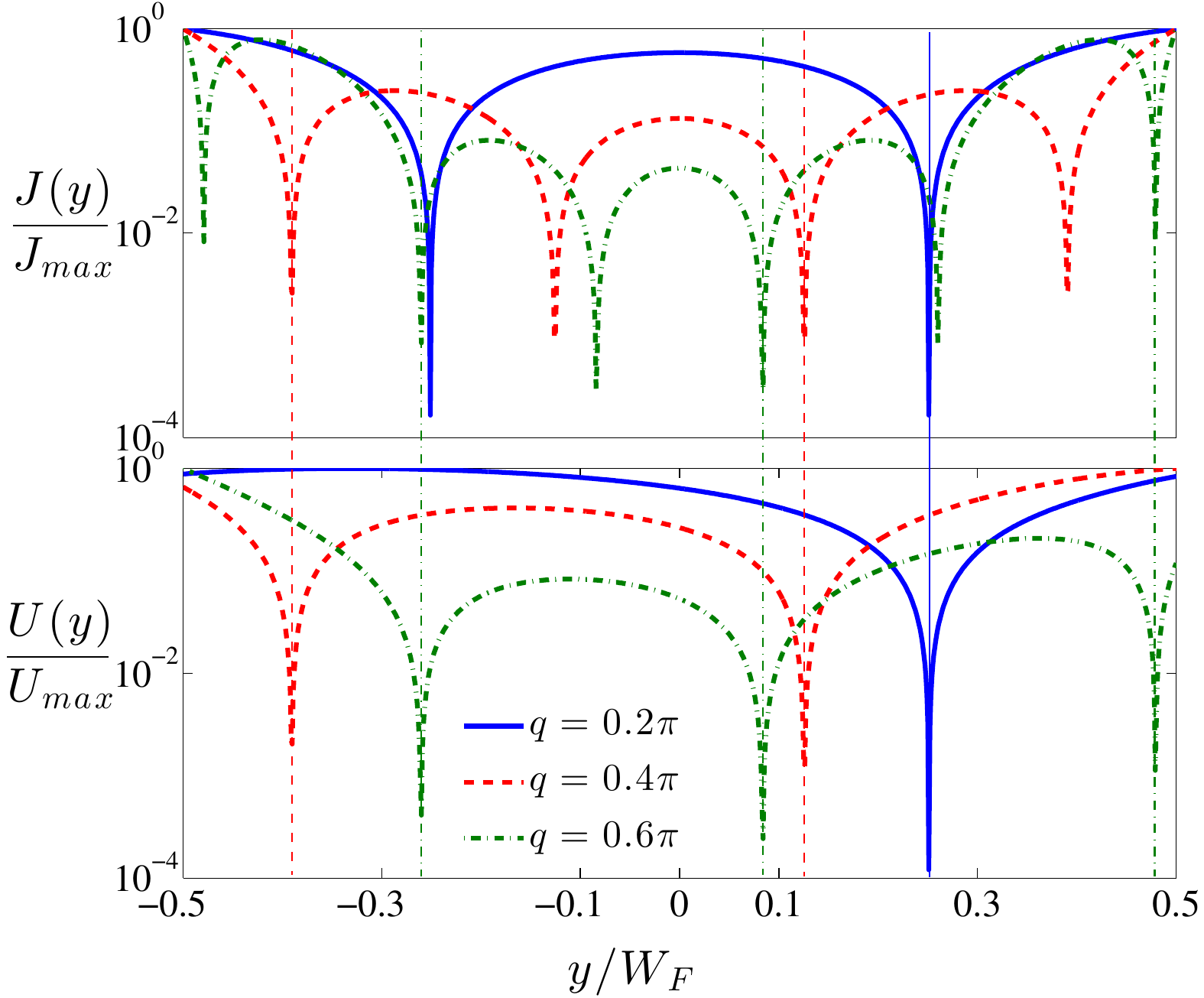}
\caption{\label{fig:vortex} (Color online). Current density $J(y)$ and
  the absolute value of the Cooper pair wave function $U(y)$,
  normalized by their maximum values, as a function of $y$ coordinate
  along the junction width at the middle of junction $x=0$ (see Fig. 1
  of main text) for three different magnetization rotation parameter $q=0.2\pi, 0.4\pi$, and $0.6\pi$. The superconducting phase difference is set fixed at $\phi=\pi/2$ and the magnetization intensity is equal to $h_0=6.0\pi\Delta$. 
}
\end{figure}

\subsection{Proximity Vortices}
We have full numerically solved the Usadel equation in the presence of the helical magnetization given by Eq. (\ref{eq:mg}). Substituting the solutions into the current density expression, Eq. (\ref{crnt2}), and the absolute value of the Cooper pair wave function below
\begin{equation}
U(x,y)\equiv-\frac{T}{8}\sum_n \big\{ f_{+} + f_{-} \big\},
\end{equation}
we determine the spatial maps of the supercurrent and Cooper pair wave function inside the surface states of TI. Our results demonstrate that the current density and Cooper pair wave function are both zero at specific locations along the junction width at $x=0$ which indicates the nucleation of proximity vortices in this class of Josephson junctions. The existence of such proximity vortices were quite recently confirmed experimentally in 3D metallic SNS junctions \cite{nphys_2015_m}. To gain better insights, we investigate the proximity vortices through the analytical expressions derived from the Usadel equation. Let us first investigate the slow rotating magnetization case which results in clearer conclusions through the analytics and then discuss the helical magnetization which demands numerics. Considering a slow rotating magnetization $q< 1$ and setting $x=0$ in the solutions (\ref{sol_chir}) we arrive at the following current density and Cooper pair wave function:
\begin{eqnarray}\label{cur_dnsy}
J_x(y) = \frac{\pi}{2e}\frac{d_F^2}{\gamma^2 R_N}  {\cal N}_{0} \sin\bigg(\phi+ \frac{2 h_0d_F}{\alpha}Qy\bigg),~~~
\end{eqnarray}
and
\begin{eqnarray}\label{cur_dnsy}
U(y) = \frac{T}{4\gamma}\sum_n\frac{|\Delta|\text{csch}(k_nd_F/2)}{\sqrt{\omega_n^2+|\Delta|^2}}\cos\bigg(\frac{\phi}{2}+\frac{h_0d_F}{\alpha}Qy\bigg).~~~~~~~
\end{eqnarray}
The current density vanishes at $y/W_F=\frac{\phi\pm n\pi}{2qh_0d_F}\alpha$ where $n=0,\pm 1, \pm2, ...$ while the zeros of $U(y)$ appear at $y/W_F=\frac{\phi\pm m\pi}{2qh_0d_F}\alpha$ where $m=\pm 1, \pm 3, ...$ . The extra zeros in the current density can be undrestood by noting the circulating form of quasiparticles' paths which cancel each other at $n=0,\pm 2, \pm 4, ...$ . Therefore, at $y/W_F=\frac{\phi\pm n\pi}{2qh_0d_F}\alpha$, $n=\pm 1, \pm 3, ...$ provided that $-1/2<y/W_F<+1/2$ the current density and $U(y)$ both vanish which determines the location of normal core of each vortex.

Next, we consider the more complicated case where the magnetization follows the pattern given in Eq. (\ref{eq:mg}). In this case we obtain the current density and $U(y)$ as follows:  
\begin{eqnarray}\label{cur_dnsy}
J_x(y) = \frac{\pi}{2e}\frac{d_F^2}{\gamma^2 R_N}  {\cal N}_y \sin\bigg(\phi+ \frac{2 h_0d_F}{\alpha}\sin Qy\bigg),~~~
\end{eqnarray}
and
\begin{eqnarray}\label{cur_dnsy}\nonumber
U(y) &=& \frac{T}{4\gamma}\sum_n\frac{|\Delta|\text{csch}(k_{n,y}d_F/2)}{\sqrt{\omega_n^2+|\Delta|^2}}\cos\bigg(\frac{\phi}{2}+\frac{h_0d_F}{\alpha}\sin Qy\bigg),\\
k_{n,y}&=&\sqrt{2|\omega_n|/D + (2h_0 \cos Qy/\alpha)^2}.
\end{eqnarray}
To determine the location of proximity vortices, we have now normalized the current density and $U(y)$ by their maximum values: $J_{max}$ and $U_{max}$. Fig. \ref{fig:vortex} exhibits the normalized current density and $U(y)$ at the middle of junction $x=0$ as a function of location along the junction width $y$ for three different rotation degrees: $q=0.2\pi, 0.4\pi,$ and $0.6\pi$.
The junction length is assumed $d_F=0.35\xi_S$, $h_0=6.0\pi\Delta$, and the superconducting phase difference $\phi=\pi/2$. As seen, the rotation degree of magnetization can change the sign of current density and induces proximity vortices in addition to the other parameters $h_0$, $d_F$, and $\phi$ which can alter the patterns. The current density shows extra zeros compared to $U(y)$ due to the cancelation process described above. The vertical lines indicate the location of proximity vortices' core where $U(y)$ and current density vanish both. Notice that a proximity vortex can move along the junction simply by modulating the phase different $\phi$. In other words, the proximity induced vortices do not necessarily reside along the junction width in a symmetric fashion with respect to $y=0$.

Finally, we note that our results on the Josephson current and proximity vortices can be also valid in the Eilenberger
limit with more sophistications in final expressions \cite{zagoskin, ykby}. These complications arise due to the
quasiballistic motion and multiple Andreev reflections at the topological insulator - superconductor
interfaces. One the other hand diffusive regime allows for highly simplifying and transparent calculations.

\section{Conclusions}\label{Conclusion}%
In conclusion, we have derived the Eilenberger and Usadel equations which describe the quasiparticles in the disordered surface states of a TI in the presence of superconductivity and magnetism. We employed this approach to study the supercurrent flow through a TI with proximity induced in-plane helical magnetization. In contrast to the case of S-TI-S junction with uniform magnetization, our results reveal that the helical magnetization can induce multiple supercurrent reversals and proximity vortices upon varying the junction thickness, magnetization strength, and helical magnetization parameters.

\section{Acknowledgments}%
 We acknowledge support from the Swiss NF and NCCR QSIT. We would like to thank A. Yu. Zyuzin for helpful discussions and G. Sewell for his valuable instructions
in the numerical parts of this work.

\end{document}